\documentstyle[preprint,prb,aps,epsf]{revtex}

\begin{document}
\baselineskip=18pt
\input{psfig}
\draft
\preprint{LA-UR-95-1639}

\title{Frustrated Spin-Peierls Chains}

\author{Jun Zang and A. R. Bishop}
\address{Theoretical Division and Center for Nonlinear Studies}
\address{Los Alamos National Laboratory, 
Los Alamos, NM 87545}

\author{D. Schmeltzer}
\address{Department of Physics, City College of CUNY, New York, N.Y. 10031}
\maketitle
\begin{abstract} 
We study the phase diagram, ground state properties, and excitation gaps
of a frustrated spin-Peierls chain, i.e. a spin-Peierls chain
with both nearest neighbor exchange $J_1$ and next nearest 
neighbor exchange $J_2$. The phase diagram is calculated
using a bosonization Renormalization-Group (RG) method.
We propose a (bosonic) bond-operator mean field 
approximation to calculate the ground state properties and excitation
gaps.
\end{abstract}


\section{Introduction}
A theoretical understanding of the frustrated spin-Peierls chains
({\it i.e.} frustrated $J_1-J_2$ Heisenberg chains with spin-lattice
interaction) is important for a number of reasons:
First, there are a number of quasi-one-dimensional systems which
can be described by a spin-Peierls model at low temperatures,
and in some of these systems, such 
as  $CuGeO_3$ materials
\cite{emery-aps} and some MX materials \cite{hro}, 
the competition between nearest
neighbor $J_1$ and next nearest neighbor $J_2$ plays an important
role. Second, the frustrated spin-Peierls model poses interesting
theoretical problems.
The frustrated $J_1-J_2$ Heisenberg model has been studied
in the literature \cite{gosh,majumdar-1/2,haldane-j,sutherland,jp}. 
Majumdar \cite{majumdar-1/2} showed that at $J_2/J_1=1/2$
the ground state is dimerized and the exact ground state wave function
is 
\begin{eqnarray}
\Psi_1 &=& [1,2][3,4] \cdots [2N-1,2N]
\nonumber \\
\Psi_2 &=& [2N,1][2,3] \cdots [2N-2,2N-1],
\label{eq:wf2}
\end{eqnarray}
where $[i,j]$ denotes the normalized singlet combination of the
spins on sites $i$ and $j$.
Later it was shown by Haldane \cite{haldane-j} that dimerization
exists at $J_2/J_1 > J_c \sim 1/6$. For $J_2/J_1 < J_c$,
the system is in a gapless spin liquid phase.
This result was confirmed by numerical studies on
finite chains \cite{jp,in}. The critical value $J_c$ found
by numerical studies \cite{jp,in} is $J_c \sim 0.3$. 
With spin-lattice interaction, there is dimerization
even without $J_2$ frustration due to the Peierls instability.
The interplay of $J_2$ frustration and spin-Peierls
mechanism is an interesting subject.

In this work, we consider the frustrated 
spin-Peierls system described by the Hamiltonian:
\begin{equation}
H=\sum (J_{1,i} \vec{S}_i\cdot\vec{S}_{i+1} +J_2 \vec{S}_i\cdot\vec{S}_{i+2}) 
 + {1 \over 2}\sum_i (M_I\dot{u}_i^2+ K u_i^2),
\label{eq:hamt}
\end{equation}
where $J_{1,i}=J_1-\beta_u u_i$.
In Sec.~II,
we recalculate the critical value $J_c$ for the 
$J_1-J_2$ Heisenberg model using Haldane's
method and obtain $J_c \sim 0.4$. We also study
the RG equations for the spin-Peierls system in the
adiabatic limit and find that the spin-lattice
interaction is always relevant and opens a finite
gap. In Sec.~III, 
we use a bond-operator mean field approximation
to calculate the dimerization magnitude, the
ground state energy and the excitation gap.
We find that this approximation can produce a
very accurate ground state energy for the
$J_1-J_2$ model for $J_2/J_1 < 0.6$, and the ground
state energy and gap calculated by this solution 
is exact at $J_2/J_1 =1/2$. Since our approximation
is based on the dimerization of the system,  we belive
our mean-field approximation should be reliable for
the spin-Peierls system. Our results are summarized
in Sec.~IV.

\section{Scaling Theory}

\subsection{Frustrated $J_1 - J_2$ Model}

Here we first discuss the frustrated $J_1 - J_2$ model studied
by Haldane \cite{haldane-j} for the $S=\frac{1}{2}$ spin system
\begin{equation}
H_s=\sum J_1 (S_i^x S_{i+1}^x +S_i^y S_{i+1}^y + \gamma S_i^z S_{i+1}^z)
      + J_2 \vec{S}_i\cdot\vec{S}_{i+2} ,
\label{eq:ham2}
\end{equation}
where the anisotropy parameter $\gamma$ is introduced.

Using the Jordan-Wigner transformation, 
Hamiltonian (\ref{eq:ham2}) transforms into
the spinless Fermion Hamiltonian
\begin{eqnarray}
H_s = \sum -\frac{J_1}{2} (c_i^{\dagger} c_{i+1} +H.C.) + J_1\gamma
(n_i-1/2)(n_{i+1}-1/2) + 
\nonumber \\
J_2{(n_i-1/2)(n_{i+2}-1/2)+[c_i^{\dagger}
(n_{i+1}-1/2)c_{i+2} + H.C.]}
\label{eq:ham3}
\end{eqnarray}

In the continuum limit, this Hamiltonian becomes
\begin{eqnarray}
H_s/(J_1 a_0)= \int dx \big(i \psi^{\dagger}_{+}\nabla\psi_{+}
-i \psi^{\dagger}_{-}\nabla\psi_{-} \big)
+g_4 (\rho_+ \rho_+ + \rho_- \rho_-) + 2g_2(\rho_+ \rho_-)
\nonumber \\
-g_3 (\psi^{\dagger}_{+}(x)\psi_-(x)\psi^{\dagger}_{+}(x+a_0)
\psi_-(x+a_0) + H.C.),
\label{eq:ham5}
\end{eqnarray}
where $g_4 = \gamma +2J_2/J_1$, $g_2 = 2\gamma + 2J_2/J_1$,
and $g_3 = \gamma - 3J_2/J_1$. Here our coefficients $g_i$ are different
from Haldane's  \cite{haldane-j}.  The crucial difference
is in the $g_3$ term. The difference
comes from the term 
$\psi^{\dagger}_{+}(x)\psi_-(x+2a_0)\psi^{\dagger}_{+}(x+a_0)\psi_-(x+a_0)$.
This term is written as $(\psi^{\dagger}_{+}\nabla\psi^{\dagger}_{+})
\psi_-\nabla\psi_-$ in Ref.\cite{haldane-j}, which has opposite
sign to the term from 
$\psi^{\dagger}_{+}(x)\psi_-(x)\psi^{\dagger}_{+}(x+2a_0)\psi_-(x+2a_0)$.
However, by direct bosonization
\cite{fradkin-bk,ha}, both expresses contribute negative Umklapp
term: $\cos(4\phi)$. Here $\phi$ is the boson phase
field:
\begin{equation}
\phi(x)=-{i\pi\over L}\sum_{p \neq 0}{1\over p}e^{-\alpha |p|/2-ipx}
[\rho_+(p)+\rho_-(p)]-(N_++N_-){\pi x\over L} ,
\label{eq:phase}
\end{equation}
where $N_{\pm}$ are the number operators.
Thus we have 
$g_3=\gamma -3J_2/J_1$ instead of $g_3=\gamma-6J_2/J_1$. 

\begin{figure}
\centerline{
\epsfxsize=10.0cm \epsfbox{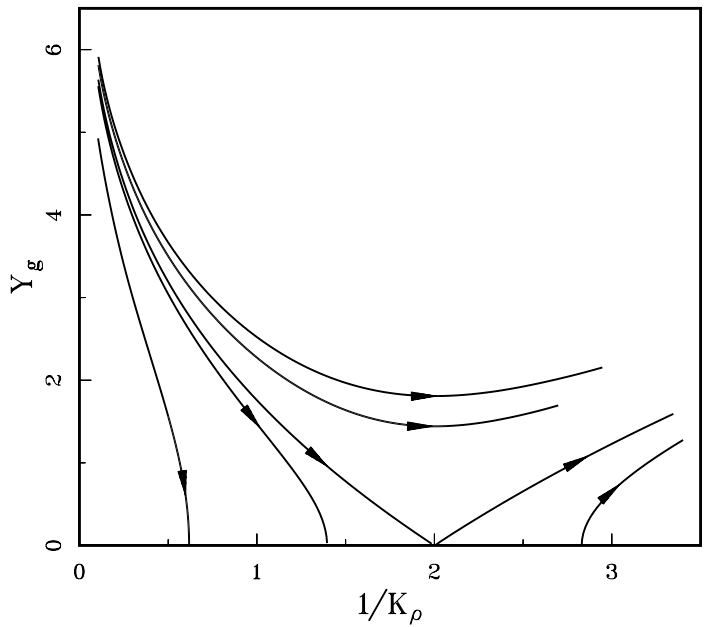}
}
\caption{Scaling Diagram for $J_1-J_2$ Model}
\label{fig:sd1}
\end{figure}

The scaling theory for model (\ref{eq:ham5}) is well known \cite{kosterlitz,jose}.
The phases for the isotropic model $\gamma \sim 1$ are identified
by Haldane \cite{haldane-j} as Neel state, gapless spin liquid, and
dimer state (with a finite gap). 
The scaling digram shown in Fig.(\ref{fig:sd1})
is calculated numerically by integrating the
scaling equations. The critical value of $J_2/J_1$
is calculated numerically by locating the intersection
of the phase boundary curve and curve 
$1/K_{\rho}=[(\pi+13/3\gamma-4/3 g_3)/(\pi-\gamma)]^{1/2}$.
The critical value we obtained is
$J_c \sim 0.447$ instead of $0.16$. The critical value
calculated by Density-Matrix-Renormalization-Group method
using 300 sites \cite{in} is $J_c=0.298$. Since we only used the lowest
order scaling equations and the critical point of $J_2/J_1$
is {\it not} close to $1/K_{\rho}=2.0$, the value of $J_c$
calculated here is not accurate. The true critical value
of $J_c$ should be close to the numerical value \cite{in,jp} $0.3$.

\subsection{Frustrated Spin-Peierls Model}

The lattice part of the Hamiltonian (\ref{eq:hamt}) 
can be written as follows (after rescaling $u$ and going
 to the continuum limit):
\begin{equation}
H_p = \int dx (1/(2\rho_I)\Pi^2_{2k_f}+1/2 k u_{2k_f}^2 + 
\frac{\gamma_u u_{2k_f}}{\pi \alpha} \cos(2\phi)),
\label{eq:hamp}
\end{equation}
where $\alpha$ is the cutoff and $\phi$ is the boson field defined
in Eq.(\ref{eq:phase}). We have use the following equation
to derive Eq.(\ref{eq:hamp}):
\begin{equation}
\psi_{\pm}(x) =\frac{1}{(2\pi\alpha)^{1/2}} \exp[\pm 2i 
\phi_{\pm} (x)]
\end{equation}
with the definition
\begin{equation}
\phi_{\pm} (x) = \frac{1}{2} \left( \phi(x) \mp \int_{-\infty}^{x} 
\partial_0 \phi(x') dx' \right) .
\end{equation}
In terms of the boson field $\phi(x)$ the total Hamiltonian is expressed
by
\begin{equation}
H=H_0+H_p+H_I ,
\label{eq:hamt0}
\end{equation}
where
\begin{eqnarray}
H_0 &=& {1\over 2\pi}\int dx \left[ (\bar{v}K_{\rho})
(\partial_{\tau}\phi)^2+({\bar{v}\over K_{\rho}})
\partial_{x}\phi)^2 \right],
\nonumber \\
H_p &=& {1\over 2}\int dx 
\left[{1\over\rho_I}\Pi^2_{2k_f}+ k u_{2k_f}^2\right] ,
\nonumber \\
H_I &=& {2g_3\over (2\pi\alpha)^2}\int dx \cos(4\phi)
+{\gamma \over \pi\alpha} \int dx \cos(2\phi) u_{2k_f}
\label{eq:hamt1}
\end{eqnarray}
and
\begin{eqnarray}
\bar{v} &=& \left[ (1+{g_4\over \pi})^2-({g_2\over \pi})^2\right]^{1/2}
\nonumber \\
K_{\rho} &=& \left[ {\pi+g_4-g_2 \over \pi+g_4+g_2} \right]^{1/2} .
\label{eq:vk}
\end{eqnarray}

With the Fermion-phonon interaction, the scaling theory is much more 
complicated. The analysis is only possible in high phonon frequency
limit $\Delta \ll \omega_{2k_f}$ or in the
adiabatic limit $\Delta \gg \omega_{2k_f}$,
where $\Delta$ is the gap in the spinless Fermion spectrum.
In this paper, we will concentrate on the adiabatic limit,
which is valid for many compounds. For completeness, we
first discuss the high phonon frequency limit.

(a) $\omega_{2k_f} \rightarrow \infty$

In this limit, the phonon degrees of freedom can be 
integrated out completely and the result is to introduce an effective
{\it attractive} electron-electron interaction:
$-\gamma^2/k/(2\pi\alpha)^2\cos(2\phi)$. Then the
phonon effect is to renormalize the Umklapp interaction:
\begin{equation}
\tilde{g}_3 = g_3 - \gamma^2/k .
\label{eq:reg3}
\end{equation}
The phase boundary between the spin liquid and
dimerized state is changed accordingly. For weak electron-phonon
coupling $\gamma^2/k \ll 1$, the dimerization
critical point changes from $J_c \rightarrow J_c-\gamma^2/k/3$.
For strong electron-phonon coupling 
$\gamma^2/k > 1$, the system is always dimerized.

(b) {\it finite high phonon frequency limit} $\omega_{2k_f} \gg \Delta$

For high phonon frequency $\omega_{2k_f} \gg \Delta$
and weak electron-phonon interactions, we can still
integrate out the phonon degree of freedom and
obtain a phonon-induced retarded attractive electron-electron
interaction \cite{fradkin-hirsch,david}:
\begin{eqnarray}
{\delta g_3\over (2\pi\alpha)^2} \cos(4\phi)
&=& -\gamma^2/k \int dt\int dt' 
\bar{\psi}(x,t)\psi(x,t)D_0(t-t')\bar{\psi}(x,t')\psi(x,t')
\nonumber \\
&=& -{\gamma^2/k\over (2\pi\alpha)^2} \cos(4\phi)
-(\gamma^2/k)\sum_{n=1}^{\infty}\left[
{\partial^n\over\partial t^n}J(x,t)\right]^2 {1\over\omega^{2n}}.
\label{eq:dg3}
\end{eqnarray}
where 
\begin{eqnarray*}
J(x,t)&=&\bar{\psi}(x,t)\psi(x,t)
\\
&=&\psi^{\dagger}_+(x,t)\psi_-(x,t)+\psi^{\dagger}_-(x,t)\psi_+(x,t)
\end{eqnarray*}
and $D_0(t)=\omega/2\exp(-i|t|\omega)$ is the bare
phonon Green's function. All the terms containing explicit
derivatives are superficially irrelevant
operators. So for finite high frequency phonons
and weak electron-phonon interaction, the retardation effects
will be scaled out at $l^* \sim v_f/\omega_{2k_f}$ or
$E_f(l^*) \sim \omega_{2k_f}$ \cite{voit-shultz}. The
critical theory is governed by the limit 
$\omega_{2k_f} \rightarrow \infty$.

(c) {\it adiabatic limit} $\omega_{2k_f}\rightarrow 0$

In the adiabatic limit, we should ``integrate'' out the
electronic degree of freedom first and then study
the effective phonon system. To this end, we can use the
following strategy to study
the spin-Peierls system in the mean-field approximation
for the phonon effect: we can first study the electronic
Hamiltonian by assuming a mean lattice dimerization
$u_{2k_f}=u_0$; then we can calculate $u_0$ by minimizing
the electron-lattice total energy. In this adiabatic
limit, the electronic Hamiltonian we want to study is:
\begin{eqnarray}
H &=& {1\over 2\pi}\int dx \left[ (\bar{v}K_{\rho})
(\partial_{\tau}\phi)^2+({\bar{v}\over K_{\rho}})
\partial_{x}\phi)^2 \right]
\nonumber \\
& + & {2g_3\over (2\pi\alpha)^2}\int dx \cos(4\phi)
+{\gamma_u u_0\over \pi\alpha} \int dx \cos(2\phi) .
\label{eq:hamad}
\end{eqnarray}

\begin{figure}[h]
\centerline{\epsfxsize=10cm \epsfbox{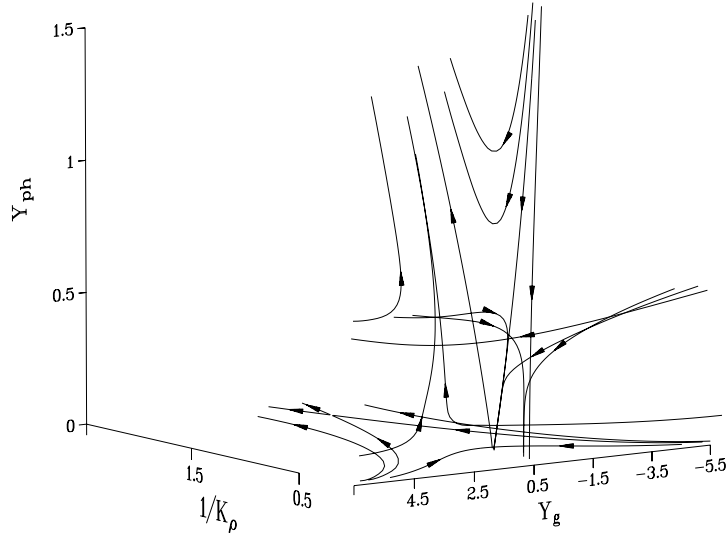}
}
\caption{Scaling Diagram
\label{fig:s3d}}
\end{figure}

In lowest order, the scaling equation can be derived
using the Coulomb gas analogy \cite{chui-lee,nelson-haperin}. 
The resulting equations are:
\begin{eqnarray}
\frac{d Y_g}{d\ln\alpha} &=& 2Y_g(1-2K_{\rho})+{1\over 2}Y_{ph}^2
\label{sc:y} \\
\frac{d Y_{ph}}{d\ln\alpha} &=& Y_{ph}(2-K_{\rho}+Y_g)
\label{sc:x} \\
\frac{d K_{\rho}}{d\ln\alpha} &=& -{K_{\rho}^2\over 2}
(Y_{ph}^2+Y_g^2) ,
\label{sc:k}
\end{eqnarray}
where $Y_g=g_3/\pi$ and $Y_{ph}=\gamma_u u_0\alpha$.
The scaling diagram Fig.(\ref{fig:s3d}) following from 
Eq.(\ref{sc:y})-(\ref{sc:k}) is calculated numerically by 
standard ordinary
differential equations integration.
In the Coulomb gas analogy, the system is equivalent to
a Coulomb gas of four kinds of particles 
({\it i.e.} two fugacities $Y_g$ and $Y_{ph}$) with charges
($\pm e_0, \pm 2e_0$). Here $\bar{\beta} e_0=(2\theta)^{-1}$. 
The activities
$\lambda_i$ of these particles are
 $\exp(-\bar{\beta}\lambda_1) = Y_g/(2\alpha)^2$ for ($\pm 2e_0$)
particles and
 $\exp(-\bar{\beta}\lambda_2) = Y_{ph}/(2\alpha)$ 
for ($\pm e_0$) particles.
The transition from the gapless phase to the gapped phase is of 
Kosterlitz-Thouless \cite{kt-tran} type. If there
is no coupling between these two fugacities, the two coupling
constants $Y_{ph}$ and $Y_g$ 
define two separate transition ``temperatures''
approximately at $|Y_g| = |1-1/\theta|$
and $|Y_{ph}| = |1-1/(4\theta)|$, corresponding to the
breaking of pairs ($-2e_0,2e_0$) and  ($-e_0,e_0$), respectively.
For each of the fugacities, the scaling trajectories are well known \cite{kosterlitz,jose}, see Fig.~\ref{fig:sd1} for the 
scaling diagram of $1/K_{\rho}-Y_g$, 
the scaling diagram of $1/K_{\rho}-Y_{ph}$ can be obtained
by rescaling $1/4$ of $1/K_{\rho}$ axis in Fig.~\ref{fig:sd1}. 
The coupling between the fagucites
contributes to the scaling equations in two respects: 
(1) $K_{\rho}$ is renormalized by $Y_{ph}$ and $Y_g$
simultaneously; see Eq.(\ref{sc:k}); (2) the close ``pairs''
with nonzero net charge will contribute to
the renormalization of $Y_{ph}$ and $Y_g$. Physically, it is
easy to see that the transition from gapless to gapped
phases is controlled by the unbinding of $(e_0,-e_0)$
pairs, since the unbinding of $(e_0,-e_0)$
pairs has a lower transition temperature. 
So the scaling process is dominated by the
$Y_{ph}$ scaling (for initial $Y_{ph} \neq 0$) (see Fig.~\ref{fig:s3d}):
the separatrix line is through the point $(Y_{ph}=0,Y_g=0,1/K_{\rho}=1/2)$.
The gapless spin liquid phase exists only at $1/K_{\rho}<1/2$.
Since for general value of $J_1$ and $J_2$, $1/K_{\rho} > 1$,
the system is always in the gapped phase.

Since we have used the bosonization procedure,
$g_i$ and $\gamma_u u_0$ are treated as small parameters.
Consequently, we can not expect that the above scaling theory is correct
for large $g_3$ or large $J_2/J_1$. Most importantly,
since we used a continuum model, ``lattice-effects'' can
not be recovered. From analytical
results \cite{majumdar-1/2} we know that in the special
limit $J_2/J_1=1/2$, the ground state is a perfect 
dimer configuration which should be true only for a discrete model.
Intuitively, we may expect that if we keep $J_2$
finite and $J_1 \rightarrow 0$, the system behaves as two decoupled
gapless spin liquids. This phase does not occur in the above
scaling phase diagram. Nevertheless, from the scaling theory,
we can know that for the $J_1-J_2$ Heisenberg model, there is
a gapped phase at $J_2/J_1 > J_c$ (but not $J_2/J_1 \rightarrow \infty$),
and from the exact results at $J_2/J_1=1/2$, the
ground state in this gapped phase is dimerized.
For the frustrated spin-Peierls model (in the adiabatic limit),
from the scaling theory, there is always a gap due to
spin lattice coupling.

\section{Bond-Operator Mean Field Solution}

From  Ref.\cite{majumdar-1/2}, we know the ground states
of the $J_1 - J_2$ model at $J_2/J_1=1/2$ are degenerate dimer states
with wave functions
\begin{eqnarray}
\Psi_1 &=& [1,2][3,4] \cdots [2N-1,2N]
\nonumber \\
\Psi_2 &=& [2N,1][2,3] \cdots [2N-2,2N-1] .
\label{eq:wf3}
\end{eqnarray}
The continuum limit discussed above showed that the gapped
state will exist at $J_2/J_1 > J_c \sim 0.3$. Thus we can expect
these gapped states to be dimer states similar to the $J_2/J_1=1/2$ case.
 Since we know that in this region the
low energy states tend to have dimer valence bonds, we will
use the bond-operator representation of the spins
first proposed by Sachdev and Bhatt for the 2D model \cite{sachdev}.

\begin{figure}
\centerline{\psfig{file=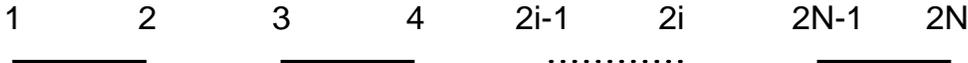,width=13cm}}
\vspace{0.4cm}
\caption{Dimerization of the Chain
\label{fig:chain}}
\end{figure}
The bond-operator representation depends on the dimerization
of the valence bond. Let us choose a dimer pattern as in Fig~{\ref{fig:chain}.
For each pair of spins connected in the dimer pattern, we can use
four bond spin operators which are defined by the singlet 
$|s\rangle$ and triplet $|t_{i\in \{1,2,3\}}\rangle$ states:
\begin{eqnarray}
|s\rangle = \hat{s}^{\dagger} |0\rangle =\frac{1}{\sqrt{2}}
(|\uparrow\downarrow\rangle-|\downarrow\uparrow\rangle)
\nonumber \\
|t_1\rangle = \hat{t}_1^{\dagger} |0\rangle =\frac{-1}{\sqrt{2}}
(|\uparrow\uparrow\rangle-|\downarrow\downarrow\rangle)
\nonumber \\
|t_2\rangle = \hat{t}_2^{\dagger} |0\rangle =\frac{i}{\sqrt{2}}
(|\uparrow\uparrow\rangle+|\downarrow\downarrow\rangle)
\nonumber \\
|t_3\rangle = \hat{t}_3^{\dagger} |0\rangle =\frac{1}{\sqrt{2}}
(|\uparrow\downarrow\rangle+|\downarrow\uparrow\rangle)
\label{eq:bop}
\end{eqnarray}

The the representations of the four bond operators $\hat{s}_n$ and $\hat{t}_{n,i\in \{1,2,3\}}$ is a faithful representation of
the original spin operators and these bond operators 
can be treated as boson operators. Their relation to the
spin operators $\vec{S}_n$ is as follows:
\begin{eqnarray}
S^{\alpha}_{2n} = \frac{1}{2} (-\hat{s}^{\dagger}_n \hat{t}_{n,\alpha}
-\hat{t}^{\dagger}_{n,\alpha}\hat{s}_n-i\epsilon_{\alpha\beta\gamma}
-\hat{t}^{\dagger}_{n,\beta}\hat{t}_{n,\gamma})
\nonumber \\
S^{\alpha}_{2n-1} = \frac{1}{2} (\hat{s}^{\dagger}_n \hat{t}_{n,\alpha}
+\hat{t}^{\dagger}_{n,\alpha}\hat{s}_n-i\epsilon_{\alpha\beta\gamma}
-\hat{t}^{\dagger}_{n,\beta}\hat{t}_{n,\gamma}) .
\label{eq:spin}
\end{eqnarray}

Using the bond operators, the Hamiltonian becomes:
\begin{equation}
H= H_0 + H_1 +H_2 +H_3,
\end{equation}
where
\begin{eqnarray}
H_0&=&\sum_n (J_1+u)(-\frac{3}{4}\hat{s}^{\dagger}_n\hat{s}_n
+\frac{1}{4} \hat{t}^{\dagger}_{n,\alpha}\hat{t}_{n,\alpha})
+\frac{1}{2}NK u^2 -\sum_n \mu (\hat{s}^{\dagger}_n\hat{s}_n
+\hat{t}^{\dagger}_{n,\alpha}\hat{t}_{n,\alpha}-1)
\nonumber \\
H_1&=&\sum_n \frac{1}{4}(-J_1+u+2J_2)
(\hat{t}^{\dagger}_{n,\alpha}\hat{t}^{\dagger}_{n+1,\alpha}
\hat{s}_n\hat{s}_{n+1}
+\hat{t}^{\dagger}_{n,\alpha}\hat{t}_{n+1,\alpha})
\hat{s}^{\dagger}_{n+1}\hat{s}_n + H.C.)
\nonumber \\
H_2&=&\sum_n \frac{1}{4}(J_1-u+2J_2)
(-\hat{t}^{\dagger}_{n,\alpha}\hat{t}^{\dagger}_{n+1,\alpha}
\hat{t}_{n+1,\beta}\hat{t}_{n,\beta}
+\hat{t}^{\dagger}_{n,\alpha}\hat{t}^{\dagger}_{n+1,\beta}
\hat{t}_{n+1,\alpha}\hat{t}_{n,\beta})
\nonumber \\
H_3&=&\frac{1}{4}(J_1-u)\epsilon_{\alpha\beta\gamma} 
(i\hat{t}^{\dagger}_{n,\alpha}\hat{t}^{\dagger}_{n+1,\beta}
\hat{t}_{n+1,\gamma}\hat{s}_n 
+i\hat{t}^{\dagger}_{n+1,\alpha}\hat{t}^{\dagger}_{n,\beta}
\hat{t}_{n,\gamma}\hat{s}_{n+1} + H.C.) .
\label{eq:ham0}
\end{eqnarray}
Here we have used a mean field approximation for  the phonon
part. Because of the $-(3/4)\hat{s}^{\dagger}_n\hat{s}_n$
term in $H_0$, it is clear that the $s$ bosons will condense:
$\langle \hat{s}_n \rangle = s_0$. The terms in $H_1$ and
$H_2$ suggest a nonzero value of 
$\langle \hat{t}_{\alpha} \hat{t}_{\alpha}\rangle$. As $J_2$ increases,
the $t_i$ bosons will also condense. Whether $t_i$
condense or not, we can omit $H_3$ in the mean field
approximation. Using this procedure,
the Hamiltonian becomes:
\begin{eqnarray}
H = N[-\frac{3}{4} (J_1 +u)s_0^2-\mu s_0^2 +\mu +Ku^2]
+\frac{1}{4} (2J_2+J_1-u) B^2 - \frac{1}{4} (2J_2+J_1-u) A^2
\nonumber \\
+
(J_2+J_1/2-u/2)A\cos k]\hat{t}^{\dagger}_{k,\alpha}\hat{t}_{k,\alpha}
+ \sum_k [\frac{1}{4}(2J_2+u-J_1) B\cos k]
\hat{t}^{\dagger}_{k,\alpha}\hat{t}^{\dagger}_{-k,\alpha}+H.C.
\nonumber \\
+\sum_k [\frac{1}{4}(J_1+u) - \mu + (J_2+u/2-J_1/2)s_0^2 \cos k .
\label{eq:mf1}
\end{eqnarray}
The self-consistent equations (at zero temperature) are:
\begin{eqnarray}
A=s_0 + \frac{3}{N} \sum_k \frac{\epsilon_k}{2E_k}\cos k
\nonumber \\
B=s_0 + \frac{3}{N} \sum_k \frac{\Delta_k}{2E_k} \cos k
\nonumber \\
s_0^2 + s_0 + \frac{3}{N} \sum_k \frac{\epsilon_k}{2E_k} =5/2
\label{eq:sfp1}
\end{eqnarray}
and
\begin{eqnarray}
\epsilon_k &=& (J_1+u)/2 +\frac{1}{4}(2J_2+u-J_1)s_0^2 \cos k
-\frac{1}{4}(2J_2+u-J_1)(A+B)+
\nonumber \\
& &\frac{1}{4}(2J_2-u+J_1)A \cos k
\nonumber \\
\Delta_k &=& \frac{1}{4} [(2J_2+J_1-u)B-(J_2+u-J_1)s_0^2] \cos k
\nonumber \\
2Ku&=&s_0^2(1-\frac{A+B}{2})+\frac{A^2-B^2}{4}-\frac{1}{4}
\nonumber \\
E_k&=&\sqrt{\epsilon_k^2-\Delta_k^2} .
\label{eq:self}
\end{eqnarray}

The self-consistent equations (\ref{eq:sfp1}) 
and (\ref{eq:self}) can be solved iteratively.
The numerical results are summarized
 in Fig.\ref{fig:bden}-\ref{fig:bdu}.
\begin{figure}
\centerline{\psfig{file=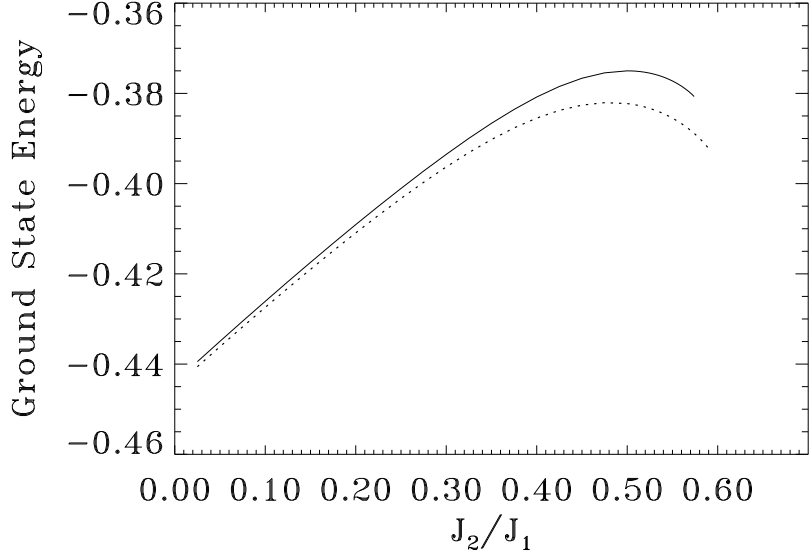,width=11cm}}
\vspace{0.4cm}
\caption{Ground state energy for $J_1-J_2$ Heisenberg model (solid
line) and frustrated spin-Peierls model (dashed line) at $K=100$.
\label{fig:bden}}
\end{figure}
\begin{figure}
\centerline{\psfig{file=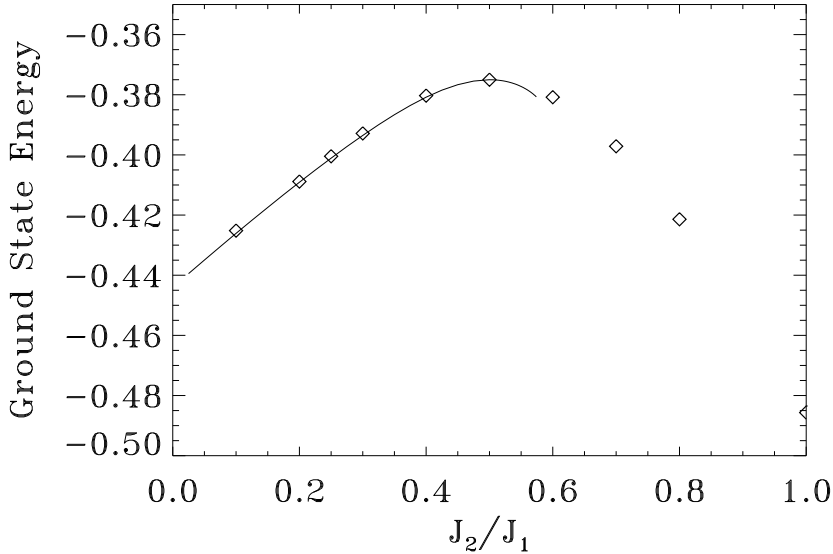,width=11cm}}
\vspace{0.4cm}
\caption{Comparison of ground state energy
from our mean-field approximation (solid line) and that
from DMRG calculation (diamond) for $J_1-J_2$ Heisenberg model.
\label{fig:cmp}}
\end{figure}
In the limit $u_0=0$, the above Hamiltonian describes the $J_1-J_2$
model studied earlier \cite{majumdar-1/2,haldane-j,jp,in}. 
We consider this case first. From equations 
(\ref{eq:mf1})-(\ref{eq:self}), it is easy to show that
at $J_2/J_1=1/2$ the ground is at $s_0=1$ and $A=B=0$,
i.e. the ground state is perfectly dimerized. The ground
state energy is $-0.375J_1$ and the gap value is $0.25J_1$, which are exact
\cite{majumdar-1/2,sutherland}. The ground state energy we calculated 
(Fig.~\ref{fig:bden})  agrees with exact numerical
results \cite{jp,in} for $J_2/J_1 \sim 0-0.59J_1$ 
(see Fig.~\ref{fig:cmp} for comparison to the calculation
in Ref.\cite{in}). This is surprising since
we do not expect dimerization for small $J_2$ without the
spin-lattice interaction. Our interpretation is that
at small $J_2$, although there is no dimerization,
the ground state is still well described by fluctuating
singlet valence bonds, so our mean-field bond
operator approximation can give a rather accurate ground
state energy. Since the spin-lattice coupling will
enhance the dimerization, we can expect that the mean field
ground state energy for the frustrated spin-Peierls model
will also be accurate.

The lattice dimerization magnitude $u_0$ for the frustrated spin-Peierls
model is shown in Fig.~\ref{fig:bdu}. Since the lattice
dimerization magnitude $u_0$  only depends on the 
ground state of the system, we believe that this mean field result
should be accurate.
From Figs.~\ref{fig:bden}\&\ref{fig:bdu} we can see that
the dimerization of the lattice is enhanced by next nearest
neighbor antiferromagnetic coupling $J_2$.
\begin{figure}
\centerline{\psfig{file=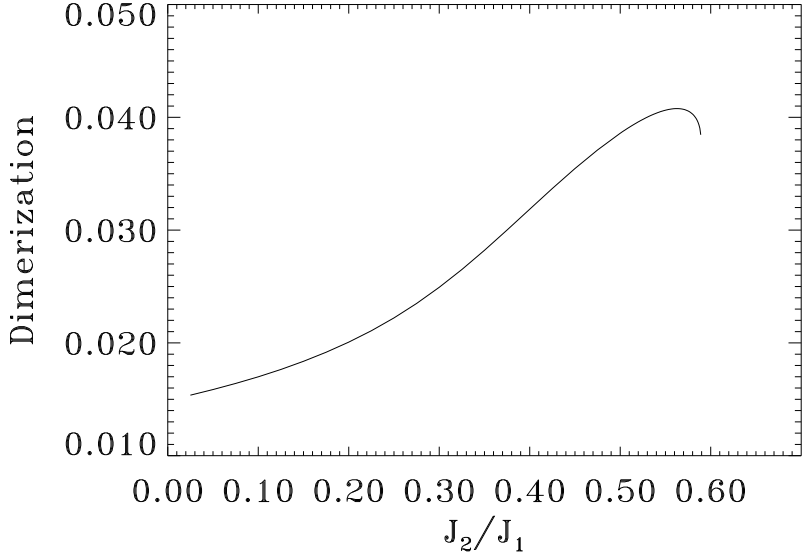,width=11cm}}
\vspace{0.4cm}
\caption{Lattice dimerization magnitude $u$ for the
frustrated spin-Peierls model  at $K=100$.
\label{fig:bdu}}
\end{figure}

The gap calculated here (Fig.~\ref{fig:bdgp}) is also reasonable
comparing to the previous results for the $J_1-J_2$ Heisenberg
model \cite{jp,in}. However there are two features of the mean-field gap 
for the $J_1-J_2$ Heisenberg model which are incorrect:
(i) We obtained a finite gap even for small $J_2/J_1$
values; and (2) The gap value starts to decrease immediately
after $J_2/J_1=1/2$ in contrast to the DMRG results \cite{in}
that the gap starts to decrease around $J_2/J_1 \sim 0.7$. 
The first incorrect feature is because we artificially assumed dimerization
in the bond-operator representation even for small $J_2/J_1$. 
For $J_2/J_1 > 1/2$, bonding between next nearest neighbors
becomes important and the description of this kind
of bonding via nearest neighbor bonding will not
be sufficient at the mean-field level. For $J_2/J_1 > 0.59$, the
$t_i$ bosons start to condense, and the gap decreases to zero.
Although the ground state energy and dimerization
due to spin-lattice interaction can be calculated
after the $t$ boson condensation, 
since we can not expect dimerization
at large $J_2/J_1$ when 
next nearest neighbor valence bond  becomes strong, 
the bond-operator representation
is questionable in this $J_2/J_1$ region. 
The bond-operator mean-field approximation should work
better for the frustrated spin-Peierls model, since even
for $J_2/J_1=0$ there is then dimerization due to Peierls
instability. Our bond-operator approximation calculation
of the gap should
be reliable for $J_2/J_1 \in [0,0.59]$.
\begin{figure}
\centerline{\psfig{file=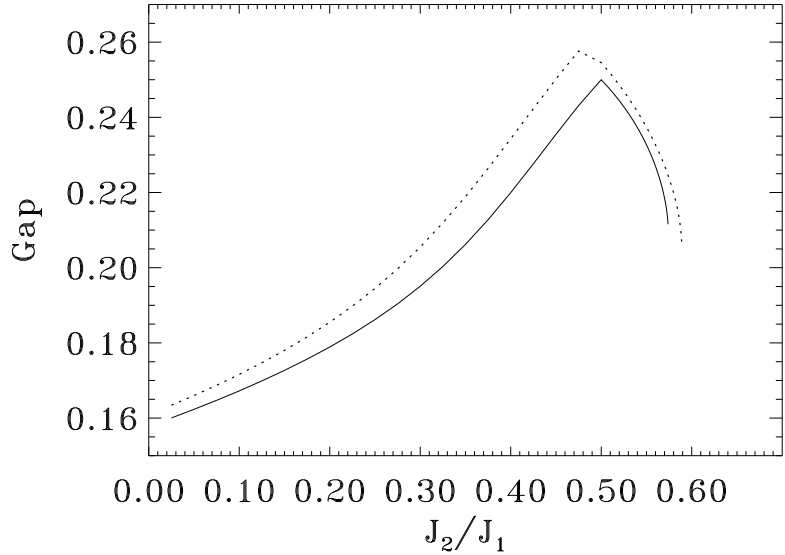,width=11cm}}
\vspace{0.4cm}
\caption{ Energy gap for the for $J_1-J_2$ Heisenberg model (solid
line) and frustrated spin-Peierls model (dashed line) at $K=100$.
\label{fig:bdgp}}
\end{figure}

\section{Summary and Acknowledgement}

In this work we have discussed the scaling behavior of a
frustrated spin-Peierls system. The ground sate energy, the
excitation gap, and lattice dimerization magnitude have been 
calculated using a bond-operator mean-field approximation.
We found that the dimerization of the spin-Peierls chain
is substantially enhanced by the next nearest neighbor
antiferromagnetic coupling $J_2$.
In the most interesting region, $J_2/J_1 < 0.59$, our bond-operator
mean field solution can describe the model rather accurately. 

This work is support by Department of Energy. DS is support
by CUNY-FRAP program.


\end{document}